# Ultra high-Q photonic crystal nanocavity design: The effect of a low-ε slab material

**Igal Bayn[*] and Joseph Salzman**

*Department of Electrical Engineering and Microelectronics Research Center, Technion Haifa, 32000, Israel*

*\*Corresponding author: eebayn@techunix.technion.ac.il*

**Abstract:** We analyze the influence of the dielectric constant of the slab on the quality factor ($Q$) in slab photonic crystal cavities with a minimized vertical losses model. The higher value of $Q$ in high-ε cavity is attributed to the lower mode frequency. The $Q$ ratio in a high-ε (silicon) vs. low-ε (diamond) slab is examined as a function of mode volume ($V_m$). The mode volume compensation technique is discussed. Finally, diamond cavity design is addressed. The analytical results are compared to 3D FDTD calculations. In a double heterostructure design, a $Q \approx 2.6 \times 10^5$ is obtained. The highest $Q \approx 1.3 \times 10^6$ with $V_m = 1.77 \times (\lambda/n)^3$ in a local width modulation design is derived.

## 1. Introduction

Slab 2D photonic crystal (PC) structures in diamond are being considered as an attractive architecture for the control and manipulation of light-matter interaction in the strong coupling regime for quantum information processing devices in the solid state [1,2]. The implementation of this architecture requires the capability to couple the optical emission of a diamond color center (NV center) to a cavity with a sharp spectral resonance. Several designs for the high-Q PC cavities on diamond were recently reported [3-5], while the best result is obtained for the double heterostructure (DH) cavity $Q \approx 7 \times 10^4$ [4]. This result is more than one order of magnitude lower than that of similar designs in silicon [6-9]. This major discrepancy in cavity quality factor ($Q$) raises the question: what is the physical mechanism responsible for the dielectric constant influence on the $Q$ value in a photonic crystal cavity?

Significant progress is made in diamond fabrication processing. The first nano-crystal diamond PC cavities were produced and optically characterized [5]. Due to high scattering the measured quality factor ($10^3$) is one order of magnitude lower than the predicted value. This demonstrates that high-Q PC realization requires single-crystalline diamond implementation, where scattering is minimized. Recently, we have reported [10,11] the first single crystalline PC fabrication for optical characterization. This process will allow realization of higher $Q$ in diamond. Thus, further effort in high-Q cavity design on diamond is justified. In addition, more deep insight in the influence of a low dielectric constant on cavity quality factor for low-ε materials is required. Thus, the main goals of the current article are to improve design for a high-$Q$, low mode volume cavity in diamond, and to uncover the physical basis for the influence of ε on $Q$.

In Section 2 we describe the analytic model of an ideal high-Q PC cavity, derived by the inverse problem method [8]. This discussion is an extension of a qualitative analysis presented in our previous work [4]. The ratio of quality factors in high-ε and low-ε materials is analyzed as a function of mode volume. The mode volume compensation technique, which allows reaching similar $Q$ for both slabs, is introduced. A simple rule for the required mode volume increase for different $\varepsilon$'s is derived. In Section 3, we present 3D-FDTD analysis of PC cavities in diamond. We start with applying the mode volume compensation to previously designed DH cavities, reaching a significant improvement in $Q$ ($Q \approx 2.6 \times 10^5$), with a mode volume $V_m \approx 1.78 \times (\lambda/n)^3$. Then, the limitations of this technique are discussed. Finally, we calculate local width modulated cavities and demonstrate the highest reported quality factor in diamond $Q \approx 1.3 \times 10^6$ with $V_m = 1.77 \times (\lambda/n)^3$. A detailed comparison of these two methods is given.

## 2. The influence of a low ε: Semi-Analytical approach

The PC cavities considered in this work are formed by a membrane suspended in air and periodically modulated by a triangular array of cylindrical air holes. In this structure the maximum bandgap is obtained between the 1$^{st}$ and the 2$^{nd}$ TE-like modes in silicon [6-9] and diamond [3-5]. Thus, we focus on the cavities with resonant frequency within this bandgap. Cavity mode is confined laterally by the Bragg reflection and vertically by total internal reflection (TIR). Therefore, cavity quality factor can be divided into two major components: $Q_l$ ($Q_v$) for the lateral (vertical) losses [6,8]. Note that $Q_l$ is determined by the number of PC periods around the cavity and its increase leads to an unbound increase in $Q_l$. Thus, by embedding the cavity in a sufficiently large PC, the value of $Q_l$ can be made arbitrarily large, and the total quality factor is limited only by $Q_v$. In the following analysis we assume that $Q=Q_v$. (This equality requires an increased number of PC periods around the cavity for the low-ε material, compared to that of the high-ε one).

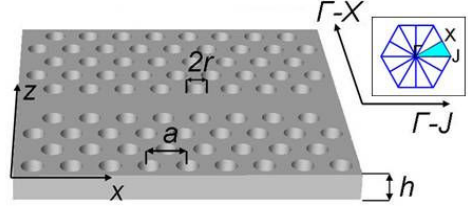

Fig. 1. The notation of the linear PC waveguide in relation to the reciprocal lattice directions. In the inset the 1st Brillouin zone of the triangular lattice (blue) and the first irreducible Brillouin zone (cyan) with high symmetry points are depicted.

*2.1 Optimized k-distribution: Ideal Gaussian Envelope cavity [8]*

We start by reviewing an analytic model for cavities formed as a spatial modulation of a PC waveguide. The PC waveguide is defined by a linear defect into the PC slab in certain direction. Here, similarly to the DH design of [7] and the local width modulated cavities [9], one row of holes in the ΓJ direction is filled (see Fig. 1). There are two waveguide modes inside the bandgap region, while we focus on the cavity design based on the lowest one with odd (even) symmetry of the $H_y$ field component along $x$ ($z$) axis in the middle of the slab ($y=0$). At $y=0$ plane, the field is defined by ($E_x$, $E_z$, $H_y$), forming TE polarization. Following the inverse problem approach, the cavity mode is represented as a product of waveguide field and a slowly varying envelope function. Thus, in $y=0$ the cavity magnetic field $H_{cav}=H_l H_w$, where the $H_l$ is an envelope and $H_w$ is a waveguide field. The vertically radiated power leaves the cavity through the light cone, which is defined in $k$-space by $k_l < k$, where $k=\omega_0/c=[k_l^2 + k_y^2]$, $\omega_0$ is cavity mode frequency, and $k_l$ lies in the $k_x k_z$ plane. Assuming that $H_y$ in the slab surface ($|y|=h/2$) can be approximated by the $H_y$ in the slab center ($y=0$), for TE-like modes, the vertically radiated power is given by [8]:

$$P_v \approx \frac{\eta}{2\lambda_0^2 k} \int_{k_l \leq k} \frac{dk_x dk_z}{k_l^2} k_z \left| FT_2(H_y) \right|^2 \tag{1}$$

where: $\eta = (\mu/\varepsilon)^{1/2}$, $\lambda_0$ is the cavity mode wavelength, $k=2\pi/\lambda_0$, $\mathbf{k_l}=(k_x\ k_z)=k(\sin\theta\cos\varphi\ \sin\theta\sin\varphi)$ and $k_y=k\cos\theta$.

The light cone is a circle in the $k$-space centered at the Γ-point (see Fig. 1 inset). In order to minimize vertical losses, $|FT_2(H_y)|$ should exhibit maxima in $k$-space as far as possible from the light cone center (Γ), and these maxima must decay fast towards this center. An ideal Gaussian field envelope satisfies both of these requirements: it is located in four J points and it decays fast towards Γ:

$$FT_2(H_y) = \sum_{k_{x_0}, k_{z_0}} sign(k_{x_0}) \exp\left(-(k_x - k_{x_0})\frac{\sigma_x}{\sqrt{2}}\right)^2 \exp\left(-(k_z - k_{z_0})\frac{\sigma_z}{\sqrt{2}}\right)^2 \tag{2}$$

where: $\sigma_x$ and $\sigma_z$ are the modal widths in real space in the $x$ and $z$ directions respectively, and ($\pm k_{x0} \pm k_{z0}$) are the J points coordinates given by $k_{x0}=\pi/a$ and $k_{z0}=2\pi/\sqrt{3}a$.

This Ideal Gaussian envelope function is regarded as an optimal and it is used to compare different designs in materials with different dielectric constant. Here we concentrate on this ideal field, while its detailed design is discussed at [8].

*2.2 Quality factor vs. high mode volume*

We assume that the PC waveguide modulation, to obtain an ideal Gaussian envelope, preserves the cavity resonance at the close vicinity to the waveguide edge at the *J* points, i.e.

$\omega_0 \approx \omega_w$, where $\omega_0$ and $\omega_w$ are the cavity and the waveguide edge frequencies. Since materials with higher $\varepsilon$ will lead to a lower $\omega_w$, we may use the following heuristic description: In a uniform slab, the frequency spectrum is $\omega = ck/\sqrt{\varepsilon_{eff}}$, and thus $\omega \propto \varepsilon_{eff}^{-1/2}$. In a PC cavity, the cavity mode $\omega_0$ is a constant and there is a full spectrum of $k$-values ($0 \leq k \leq 2\pi/a$), but still, in a similar way as in the uniform slab, $\omega_w$ and $\omega_0$ decrease with increasing $\varepsilon$. Thus, the $k$ components that are able to escape from the cavity (according to Snell's law) relate to the value of the material dielectric constant. Now, if we *assume* a Gaussian distribution in $k$-space around the $J$ points (Eq. (2)) the amplitude of the $H_y$ field with $k$ components departing from $J$ decrease exponentially, and by crossing the boundary determined by the light cone ($k_l < k$), will cause vertical losses. The magnitude of components crossing this boundary, thus depends *exponentially* on $\omega_0$, which is related to the material $\varepsilon$. As a result, the vertical losses in a slab with a high-$\varepsilon$ are substantially lower than those in a slab with a low-$\varepsilon$, which explains the different $Q$ factors.

As an example of the $Q$ values for a similar cavity design in high-$\varepsilon$ and low-$\varepsilon$ materials, we present a calculation of the ratio between vertical quality factors in Silicon and diamond ($Q_{vSi}/Q_{vD}$). The influence of the modal width in the $x$ direction ($\sigma_x$) on this ratio is examined. Note that the inverse problem design [8] enables an equal mode-width along the waveguide direction for both high and low-$\varepsilon$ materials. Therefore, in the analysis of $Q_v$ ratio in silicon and diamond cavities, we first assume: $\sigma_{xD} = \sigma_{xSi}$. The mode widths in the $z$ direction is defined by the PC bandgap confinement, and is obtained from the Gaussian fit to the best reported DH cavities in diamond [4] and in silicon [5]. This fit results are $\sigma_{zD} = 1a$ in diamond, and $\sigma_{zSi} = 0.85a$ in silicon.

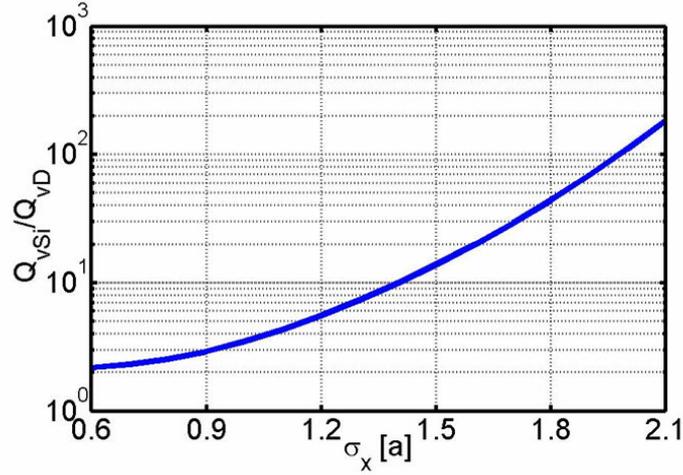

Fig. 2. The mode width $\sigma_x$ influence in silicon and diamond: $Q_{vSi}/Q_{vD}$ vs. $\sigma_x$, while $V_m \sim \sigma_x$.

In Fig. 2 the vertical quality factor ratio ($Q_{vSi}/Q_{vD}$) versus mode width in x direction is presented. The mode-width increase results in further localization of field distribution in the $k$-space, thus decreasing vertical power losses through the light cone. Since the vertical power in $k$-space decreases nearly as a Gaussian from the J points towards $\Gamma$, a similar increase in $Q_v$ as a function of $\sigma_x$ is expected for both silicon and diamond cavities. The fast increase in $Q_{vSi}/Q_{vD}$ displayed in Fig. 2 is understood since this ratio reflects the difference of $|FT_2(H_y)|$ at the light cone edges in both materials. Qualitatively, the ratio of radiation losses will behave as $exp[(k_{Si}^2 - k_D^2)\sigma_x^2]$, and increase nearly exponentially with the mode volume $V_m \propto \sigma_x$.

*2.3 Mode volume compensation technique*

As we have shown, for cavities with the same mode volume, there is a strong dependence of $Q_{vSi}/Q_{vD}$ on $\sigma_x$. In contrast, by slightly enlarging the mode volume of the cavity in the low-$\varepsilon$ material, the higher vertical losses can be compensated by a better localization in the *k*-space: for any cavity design in a high-$\varepsilon$ material, a similar design in a lower-$\varepsilon$ material can be compensated to produce a similar *Q*-value by a modest increase in the mode volume ($V_m \propto \sigma_x$).

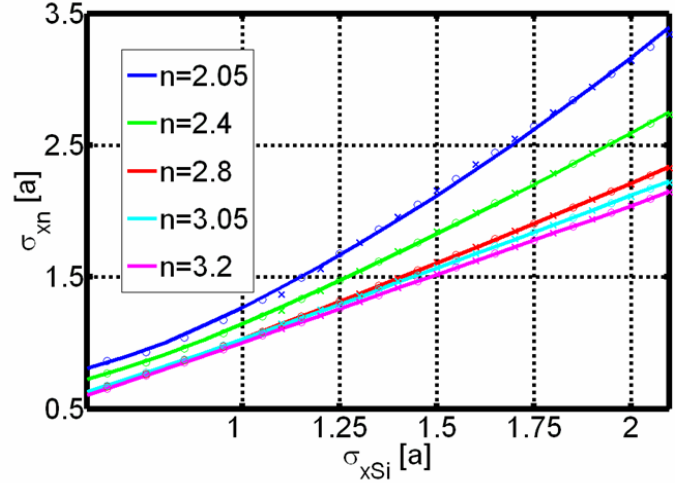

Fig. 3. Mode volume compensation: $\sigma_{xn}$ vs. $\sigma_{xSi}$, $\sigma_{xn}$ is mode width along *x* direction for slab refractive index *n*, that provides $Q_{vSi}(\sigma_{xSi})/Q_{vn}(\sigma_{xn}) \approx 1$ and $\sigma_{xSi}$ is mode width in silicon. In solid lines the calculated $\sigma_{xn}$ is shown. Hollow circles denote the analytic fit of the form $\sigma_{xn}^2 = A + B(\sigma_{xSi} - C)^2$, where A, B and C are functions of $\varepsilon = n^2$. The *x* marks denote linear fit of the form $\sigma_{xn} \approx \alpha \sigma_{xSi} + \beta$. The green line is diamond (*n=2.4*).

This compensation technique is demonstrated in Fig. 3, where we present the mode-width $\sigma_{xn}$ of the slab PC waveguide based cavity with refractive index *n*, *versus* the mode-width of a Silicon based cavity ($\sigma_{xSi}$), that provides $Q_{vSi}(\sigma_{xSi})/Q_{vn}(\sigma_{xn})=1$. These calculations are based on the characteristic mode frequencies and widths of DH cavities with refractive index *n*. One can observe, that there is an accurate fit to this mode width dependence, which is given *by* $\sigma_{xn}^2 = A + B(\sigma_{xSi} - C)^2$, where *A, B,* and *C* are constants for each *n*. A qualitative explanation of this fit follows similar arguments to these (given above) of the exponential dependence of $Q_{vSi}/Q_{vD}$. Moreover, for $\sigma_{xSi} > 0.9a$, we can approximate: $\sigma_{xn} \approx \alpha \sigma_{xSi} + \beta$, with the coefficients $\alpha$ and $\beta$ depending on $\varepsilon$ only, since $\varepsilon$ dictates the photonic band gap confinement, which defines $\sigma_z$ and mode frequency in the waveguide based cavity geometry $\omega_0$.

Following Fig. 3, for diamond PC (green line), a simple linear (approximated) relation is obtained: $\sigma_{xD} \approx 1.57\sigma_{xSi} - 0.51$. This *phenomenological* result is important since it allows us to evaluate the increase in mode volume required in order to obtain a diamond PC cavity with the same vertical quality factors as one with a similar design in silicon. As an upper bound we can state that for the same $Q_v$, the mode width in the x direction for the diamond cavity ($\sigma_{xD}$) should be ~*1.6* times larger than in silicon ($\sigma_{xSi}$). Taking into account the mode width in z direction, the upper bound for the mode volume in diamond (to obtain the same $Q_v$-value) is nearly ~*1.9* times higher than in silicon!

## 3. Waveguide based PC cavities in Diamond

In this section we discuss the mode compensation technique and its limitations by analyzing two diamond PC cavity designs: the modified double heterostructure [7] and the waveguide local width modulation [9]. Several modifications to the original structures are introduced to accommodate for the diamond low dielectric constants. The common feature of these two designs is that they both rely on a local dielectric constant modulation in the vicinity of PC waveguide.

Cavity modes were calculated with 3D-FDTD, whereas for quality factors and resonant frequencies, the Pade approximation technique is used [23]. The computation cell consists of *24×16* PC periods in *xz* plane, with the height of *2a* in the *y* direction. This cell is surrounded by the PML and mirror boundaries are applied in all directions. The calculations were initially performed with the discretization of *0.04a* [13], while a higher discretization of *0.03a* exhibited similar *Q*. The results convergence for a larger structure exhibited similar *Q*.

*4.1 Modified Double Heterostructure (DH) design*

This structure is formed by introducing two different lattice constants: $a_1$, $a_{21}$ elongated in the *x*-direction ($a_1 > a_{21} > a$) into the one missing-hole waveguide. The lattice constant in the *z* direction is unchanged. In this way, waveguide confinement is obtained in the *x* direction and band-gap confinement in the *z* direction. The number of elongated periods for *x>0* is $N_1$ and $N_{21}$ for $a_1$, $a_{21}$ respectively (see Fig. 4(a)). The detailed cavity geometry is summarized in Table 1.

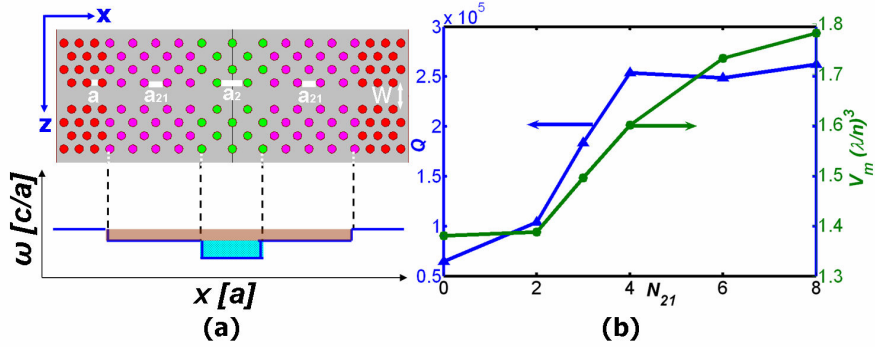

Fig. 4. Modified Double Heterostructure cavity: (a) Geometrical parameters for $N_{21}=4$ cavity, with schematic of waveguide confinement dawn below. (b) $Q$, $V_m$ vs. $N_{21}$.

Table 1. Modified/simple Double Heterostructure cavity parameters.

| *h [a]* | *r [a]* | *W [a]* | $a_1$ *[a]* | $a_{21}$ *[a]* | $N_1$ | $N_{21}$ |
|---|---|---|---|---|---|---|
| 0.96 | 0.275 | 0.977√3 | 0.05 | 0.025 | 1 | 0÷8 |

Here, the mode volume compensation is implemented by varying $N_{21}$. As one can observe in Fig. 4(b), the *1.2* times increase in the $V_m$ results in the *4* times higher *Q* ($Q(N_{21}=0\rightarrow 4)=6.4\times 10^4 \rightarrow 2.53\times 10^5$). Further increase in $N_{21}$ exhibits saturation in the increase of *Q* ($Q(N_{21}=8)\approx 2.6\times 10^5$), regardless of the higher $V_m$. This behavior is not predicted by the (approximated) analysis given in Section 2. The difference between the DH and the Ideal Gaussian mode distribution is responsible for it (along *x* axis the fit between them is nearly perfect, while for *z≠0*, it is not). As a result, the $H_y$ distribution in the $k_z \neq 0$, results in significant vertical losses, even for the increased mode volume cavities ($N_{21}=4\div 8$). These losses prevent further improvement in *Q* with the DH design.

*4.2 Local width modulation design*

This cavity is also formed by local index variation in the vicinity of the one missing-hole waveguide (*W1*) [9]. Here, adjacent holes are *z*-shifted away from the waveguide. These holes are divided in 3 groups: *A, B, C* and each is shifted by $D_A$, $D_B$, $D_C$, respectively (see Fig. 5

(a,b)). In this way a more gentle confinement is obtained in both x and z directions. The hole's radii are $r_A=0.2875a$, $r_{B,C}=r=0.275a$. The slab thickness is $h=0.96a$.

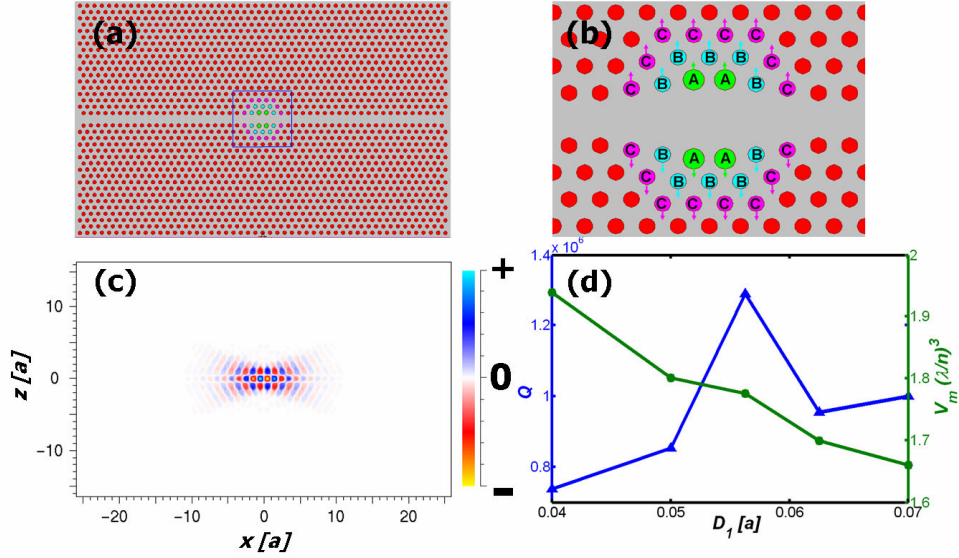

Fig. 5. Local width modulation cavities (A1): (a) A1 cavity computational domain. (b) Detailed geometry. The holes $A$, $B$, $C$ are shifted by $D_A$, $D_B$, $D_C$. (c) $H_y$ in the plane $y=0$ (d) $Q$, $V_m$ vs. $D_A$. The detailed holes shifts are summarized in Table 2.

Table 2. Local width modulation summary for the type A1 cavities in $a$ units.

| Cavity # | $D_A$ | $D_B$ | $D_C$ |
|---|---|---|---|
| 1 | 0.04 | 0.026 | 0.013 |
| 2 | 0.05 | 0.03 | 0.016 |
| 3 | 0.056 | 0.031 | 0.019 |
| 4 | 0.062 | 0.042 | 0.021 |
| 5 | 0.07 | 0.047 | 0.023 |

The cavity $Q$ is optimized by the various hole's shifts $D_A$, $D_B$, $D_C$. The results are shown in Fig. 5(d), while detailed cavity geometry is specified in Table 2. The best quality factor $Q\approx1.3\times10^6$ is obtained for $D_A=0.056a$, $D_B=0.031a$, $D_C=0.019a$, and exhibit $V_m=1.775\times(\lambda/n)^3$. It's magnetic field profile ($H_y$) at the plane $y=0$ is shown in Fig. 5(c). This is an optimal value, while for higher and lower mode volumes the values of $Q$ are lower.

On a first glance, based on the Ideal Gaussian (approximated) model, the higher $V_m$, the better $Q$ is expected. Since the increase in $V_m$ subsequently delocalize the mode in space via the increase in the mode widths in $x$ and $z$ directions, the compression in $k$-space results in a decrease in vertical power losses, and a higher $Q$ is obtained. In the width modulation design, $V_m$ deviation from its optimal causes a subsequent increase in the vertical power losses that degrade the $Q$. One can observe this behavior from the $|H_y|$ distributions presented in Fig. 6. A detailed inspection of Fig.6 reveals that any departure from the "optimal" obtained point (central part in Fig.6), bring higher intensity peaks of the field within the light cone.

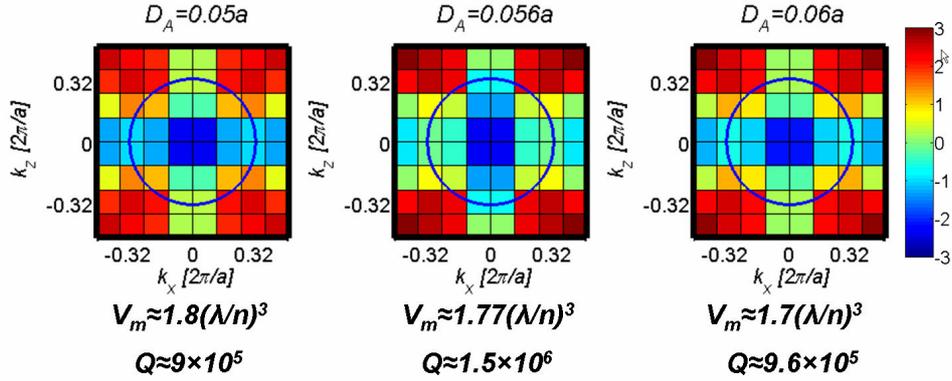

Fig. 6. $log(|FT(Hy)|^2)$ in a.u. for the A1 cavities with different $D_A$ shifts (see Table. 2).

We would like to stress that further optimization of the cavity design is possible. For instance, a gentle modulation to the $D_C$. However, this further effort, considering the operating wavelength of the diamond cavity, which is intended to be at $\lambda \sim 637.3 nm$, would require sub nanometer fabrication process resolution, which is currently beyond experimental access.

## 5 Conclusions

We have analyzed the influence of varying the material dielectric constant on PC cavity $Q$, by applying an ideal Gaussian field model [8]. The physical mechanism behind the reduced $Q$ in the waveguide based PC cavities in low dielectric constant material is explained in the terms of the mode frequency. The influence of mode volume increase via mode width elongation is given and a mode compensation technique is described. Simple analytical rules for the required mode volume compensation are derived.

In diamond, the mode volume compensation technique is applied to the modified double heterostructure deign. The best value of $Q \approx 2.6 \times 10^5$ with $V_m \approx 1.8 \times (\lambda/n)^3$ is obtained, while further improvement via mode compensation is impossible due to the $Q$ saturation. We have shown that with a modified local width cavity design the $Q \approx 1.3 \times 10^6$ with $V_m = 1.775 \times (\lambda/n)^3$ is obtained. As far as we know, this is the best ultra-high Q cavity design obtained in diamond. Moreover, this is a "near optimal" design, which exhibits the best $k$-space distribution. Further work in a diamond high-Q cavity design might improve the $Q$, but will require sub-nanometer precision in the device fabrication.

**Acknowledgement**

Igal Bayn would like to thank to Dr. Anne Weill-Zrahia, project manager of the Technion NANCO Cluster for her devotion and indispensable help that made this work possible. Support from the Russell Berrie Nanotechnology Institute is acknowledged.